\documentclass[aps,twocolumn,epsfig,graphics,showpacs,floatfix,mathbbm]{revtex4}

\usepackage{amsmath,amsfonts,amssymb,graphics,graphicx,epsfig,color,times,bbm}

\newcommand{\me}{\mathrm{e}}
\newcommand{\mi}{\mathrm{i}}

\begin{document}

\title{Inhomogeneous atomic Bose-Fermi mixtures
in cubic lattices}
\author{
M.\ Cramer$^{1}$, J.\ Eisert$^{1,2}$, and
F.\ Illuminati$^{3}$}
\affiliation{1 Institut f{\"u}r Physik, Universit{\"a}t Potsdam,
Am Neuen Palais 10, D-14469 Potsdam, Germany\\
2 QOLS, Blackett Laboratory, Imperial College London,
Prince Consort Road, London SW7 2BW, UK\\
3  Dipartimento di Fisica ``E. R. Caianiello'', Universit\`a di Salerno, 
INFN Sezione di Napoli - Gruppo collegato di Salerno, INFM UdR di Salerno, 
84081 Baronissi (SA), Italy}
\date{May 19, 2004}

\begin{abstract}
We determine the ground state properties of 
inhomogeneous mixtures of bosons and fermions 
in cubic lattices by studying the
Bose-Fermi Hubbard model including 
parabolic confining potentials.
We present the exact solution in the limit 
of vanishing hopping (ultradeep lattices) and
study the resulting domain structure of composite 
particles. For finite hopping
we determine the domain boundaries 
between Mott-insulator plateaux and hopping-dominated regions for lattices
of arbitrary dimensionality within perturbation theory.
The results are compared with a new numerical method that is
based on a Gutzwiller variational approach for the 
bosons and an exact treatment for the fermions.
The findings can be applied as a guideline for future
experiments with trapped atomic Bose-Fermi mixtures 
in optical lattices.

\end{abstract}

\maketitle

Mixtures of ultracold
bosonic and fermionic  
particles have attracted
a considerable amount of attention in recent years, to
a high extent triggered by the perspective of  
achieving prima facie transitions to superfluidity
in systems of neutral fermionic atoms \cite{Something}.
Spectacular progress has already been achieved
in the experimental manipulation of cold atoms in optical
lattices with an astonishingly high 
degree of control.
Most prominently, the superfluid-Mott insulator phase transition  
in systems of bosonic atoms in optical lattices has been 
experimentally observed \cite{Greiner}, and the
production of degenerate Fermi gases in optical lattices has
been recently achieved \cite{Lattice}.
A perspective of key interest in this field lies in the possibility of discovering and probing
new quantum phases of matter by combining  
ideas from the study of Bose-Fermi mixtures in solid state systems
and of cold quantum gases in optical lattices \cite{Ours,Maciek,Maciek2,Blatter, Keith}.
In Ref.\ \cite{Ours}, the
Bose-Fermi Hubbard (BFH) Hamiltonian
has been introduced and 
derived from the underlying microscopic many-body Hamiltonian,
linking the experimentally accessible quantities 
to the model parameters.
A mean field argument has been presented for the onset of a bosonic superfluid
transition, and in a numerical analysis the behavior of
on-site quantities in several situations has been studied.
In Refs.\ \cite{Maciek, Maciek2} the phase diagram of
homogeneous boson-fermion mixtures in optical lattices has been 
studied in a mean field approach, and the existence of a
complex structure of phases of composite fermionic
particles has been suggested.
In Ref.\ \cite{Blatter} stable
supersolid phases have been predicted for homogeneous
Bose-Fermi mixtures in two-dimensional lattices. 
Finally, Ref.\ \cite{Keith}
addresses the task of assessing the phase diagram of 
the BFH model using an exact diagonalization approach
for systems of small size.

The investigations in Refs.\ \cite{Maciek, Maciek2, Blatter, Keith} are confined 
to the homogeneous case, i.e., to a translationally invariance BFH Hamiltonian. 
%
While this is a very 
reasonable approach as far as the discussion of the actual phases 
in the thermodynamical limit is concerned, it is  
not quite the one encountered experimentally, e.g.
in the case of trapped, ultracold atomic gases. The external 
confining potential, superimposed to the optical lattice potential, 
breaks the translational symmetry, and leads to profound modifications
of the phase structure by allowing for the appearance
of spatial domains of coexisting different phases along the lattice,
as recently studied for pure bosonic \cite{Batrouni} and pure fermionic systems \cite{Batrouni2}.
Studies of such inhomogeneous systems are of 
immediate relevance for the interpretation of experimental 
findings, where some confinement in a trap is necessary.

In the present paper we study the effects of an inhomogeneous
confining potential on Mott and superfluid regions emerging in systems of 
Bose-Fermi mixtures in regular lattices at zero temperature. 
We show that the model is exactly solvable
in the limit of very strong lattices (vanishing bosonic and
fermionic hopping), and analyze the related structure of domains
of composite particles.
We then consider the general case of finite hopping
in $D$-dimensional lattices, study the bulk properties
of the system in Landau theory and local density approximation (LDA), and
determine the general phase boundaries of the different domains.
Notably, we introduce a method to treat the bosons within a Gutzwiller-type
ansatz \cite{Krauth,Jaksch} and the fermions exactly, a versatile method that is applicable
to several systems of this kind. 
This method allows us to present for the
first time the domain structure of inhomogeneous atomic mixtures in confining potentials
and the respective phase diagrams for the homogeneous case.
%

Starting point of our analysis is the single-band BFH Hamiltonian \cite{single-band},
which captures the essential properties of dilute mixtures in optical lattices
under fairly general assumptions on the tunable
physical parameters \cite{Ours}.
The grand canonical BFH Hamiltonian reads
\begin{eqnarray}
\hat{H}&=&\label{bfh}
	-J_B
	\sum_{\langle i,j\rangle}
	\left(
	\hat{b}_i^\dagger\hat{b}_j+ \hat{b}_j^\dagger \hat{b}_i\right)
	-J_F
	\sum_{\langle i,j\rangle}
	\left(
	\hat{f}_i^\dagger\hat{f}_j+
	\hat{f}_j^\dagger\hat{f}_i \right)
	\nonumber \\
	&+&U_{BB}\sum_i\hat{n}^i_B(\hat{n}^i_B-1)
	+U_{BF}\sum_i
	\hat{n}^i_B\hat{n}^i_F  \\
	&+& \sum_i
	\hat {n}^i_BV_B^i
	+\sum_i\hat {n}^i_FV_F^i
	-\mu_B\sum_i \hat{n}^i_B
	-\mu_F\sum_i
	\hat{n}^i_F \, . \nonumber 
\end{eqnarray}
Here, $\hat b_i$ and
$\hat f_i$ are the on-site bosonic and fermionic 
annihilation operators, respectively, 
whereas 
$\hat{n}^i_B=\hat {b}_i^\dagger\hat {b}_i$ and
$\hat {n}^i_F=\hat {f}_i^\dagger
\hat{f}_i$.
Sites are associated with a cubic $D$-dimensional lattice with 
fixed spacing, and $i=(i_1,...,i_D)$ 
denotes a $D$-tuple labeling the coordinates
of a site $i$ with coordination number $d$ (i.e., the number of nearest
neighbors).
The symbol $\langle i,j\rangle$ denotes summation over pairs
of nearest neighbors.
The first two terms in Eq.\ (\ref{bfh}) describe independent bosonic
and fermionic nearest-neighbor hopping with positive amplitudes
$J_B$ and $J_F$. The subsequent 
line represents on-site boson-boson and 
boson-fermion interactions.
Finally, the first two terms of the 
last line incorporate the external confining 
potential, which, in typical experimental situations, can
be taken to be harmonic.
The origin of the lattice is 
chosen to be at the minimum of the trapping potential, assumed to
be equal for bosons and fermions, 
so that 
$V_B^i = V_F^i = V_i := V_0  | i |^2$.
This Hamiltonian is a generalization to systems of bosons and fermions
of the frequently employed Bose-Hubbard model
exhibiting the Mott to superfluid phase transition in bosonic systems 
\cite{Fisher,Sachdev,Jaksch}.
Expressions linking the model 
parameters to quantities that can be tuned in an actual experimental
situation, such as the depth of the optical lattice and the atomic
scattering lengths, are provided in Ref.\ \cite{Ours}.

\begin{figure}
\includegraphics[width=\columnwidth]{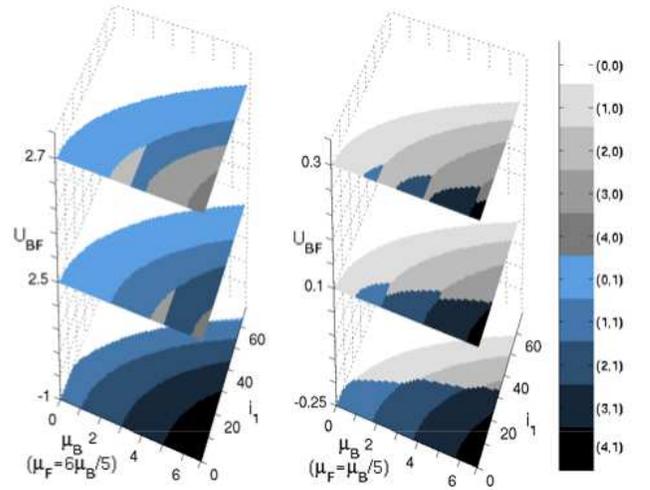}
\caption{Distribution of 
integer boson and fermion numbers for the case
$J_B=J_F=0$ and $V_0=0.002$ 
for a $D$-dimensional cubic lattice. This is  
encoded in the color as shown in the bar on the right hand side
(number of bosons, number of fermions) 
as a function of the component $i_1$,
the chemical potential $\mu_B$, and $U_{BF}$.
For the left (right) figure,
$\mu_F= 6 \mu_B/5$ ($\mu_F=\mu_B/5$) is chosen.
\label{1}}
\end{figure}

{\it I) Exact solution with vanishing hopping.} --
A surprisingly rich situation is already encountered in the case
of vanishing hopping: $J_B=J_F=0$. In this case the 
Hamiltonian $\hat{H}_0$	is simply 
a sum of single-site contributions, and the
eigenstates of the BFH model are tensor products
of number states  with state vectors
	$
	|\psi\rangle=
	|n_0, n_1, \cdots \rangle
	|m_0, m_1, \cdots\rangle$,
where $n_i=0,1,2,... $ and 
$m_i=0,1$ 
represent the occupation number of bosons and fermions
at site $i$, respectively. For simplicity of notation,
we will fix the energy scale by setting $U_{BB}=1$.
We have
$
\langle\psi |\hat{H}_0| \psi \rangle=
	\sum_i (
	n_i^2-n_i+{U}_{BF}n_im_i+
	{V}_i (n_i + m_i)-{\mu}_Bn_i-{\mu}_Fm_i)= :
	\sum_i E(n_i,m_i)$,
where for the ground state with state vector $|\psi_0\rangle$
the occupation numbers take the specific values
\begin{eqnarray*} 	
	\bar n_i=
	\left\{
	\begin{array}{l}
	\max (0,[(1+{\mu}_B-{V}_i)/2]), \text{ if }
		E(\bar n_i,0)<E(\bar n_i,1),\\
		\max (0,[(1+ {\mu}_B -
  {V}_i - {U}_{BF})/2]), 
 \text{ otherwise},
	\end{array} \right.& &\\
     \bar m_i =
    \left\{
    \begin{array}{ll}
    0,   \text{ if } E(\bar n_i,0)<E(\bar n_i,1), \\
    1,   \text{ otherwise},
     \end{array}\right.\hspace{1cm}& &
\end{eqnarray*} 
where $[.]$ denotes the closest integer to the value in brackets.
According to the above determination, several types of composite 
particles can be formed.
Composites consisting of $\bar m_i$ fermions and $\bar n_i$ bosons 
are formed at site $i$, see Fig.~\ref{1}. Connected domains
with fixed integer particle number are formed and, depending on the interaction
strength $U_{BF}$ and the relation of the respective
chemical potentials $\mu_B$ and $\mu_F$, 
the fermions distribute around the center of
the trap or are pushed outwards.

\begin{figure}
\includegraphics[width=\columnwidth]{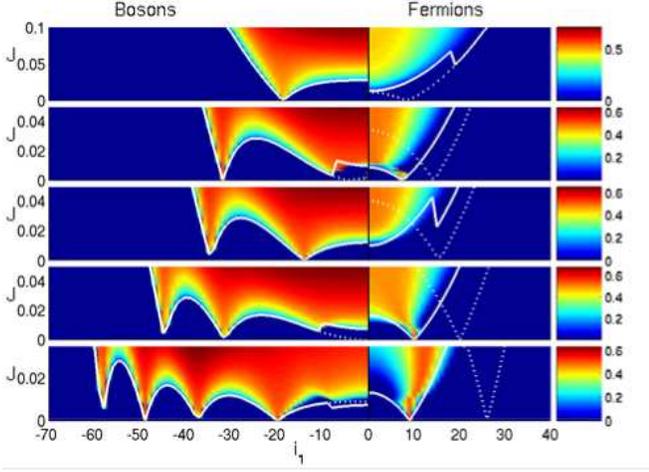}
\caption{The boundaries between
Mott and hopping-dominated regions for $D=3$ ($d=6$) 
for both bosons (left) and fermions (right), as a function of
the site index $i_1$, and of the hopping $J=J_B=J_F$. 
The values of $U_{BF}=0.3$ and $\mu_F=\mu_B/5$ 
are chosen (corresponding
to the topmost plot on the right of Fig.\ \ref{1}), 
and, from top to bottom,
$\mu_B=(0.7, 2, 2.4, 4, 6.8)$. In each plot,
the $J=0$ axis corresponds to the plots of
Fig.\ \ref{1}. The white solid 
line depicts the phase boundaries as
determined in section II. The dashed line reproduces the same 
plots, but for $U_{BF}=0$.
In the same diagram, the background 
color encodes the variance of the on-site
density 
$\sigma_{B/F}^i=(\langle (\hat{n}^i_{B/F})^2\rangle-\langle{\hat{n}^i_{B/F}}\rangle^2)^{1/2}$ 
from the numerical variational analysis discussed in section III. Dark blue (gray) 
corresponds to the Mott region with $\sigma_{B/F}^i=0$.
}\label{2}
\end{figure}

{\it II) Finite hopping: perturbative treatment.} -- 
We now turn to the strong coupling limit
where also small but finite hopping is allowed for. 
In a wide range of physical parameters,
the strength of the hopping 
for bosons and fermions are approximately
of the same value, for instance for neutral atoms 
in optical lattices \cite{Ours}. We set subsequently 
$J_F=J_B=:J$, and treat the small positive
parameter $J$ as a perturbation. As in Refs.\ 
\cite{Sheshadri},
we introduce a mean field approximation, which 
amounts to a replacement of the bosonic operator products
in Eq.\ (\ref{bfh}) according to
$\hat{b}_i^\dagger \hat{b}_j  \longmapsto  {\psi_B^i}^* \hat{b}_j +
\hat{b}_i^\dagger{\psi_B^j} - {\psi_B^j} {\psi_B^i}^*$,
the complex numbers
$\psi_B^i$ being variational parameters modeling the influence
of neighboring atoms with the physical interpretation of a superfluid
parameter. 
We consider the resulting corrections 
to the ground state energy, $\langle\psi_0 |\hat{H}_0| \psi_0 \rangle=
\sum_iE(\bar n_i,\bar m_i)$, to second order in  $J$. 
Moreover,
to study bulk properties we will make use of the local density
approximation (LDA). This means taking for each lattice site 
$\psi_B^i$ to be equal to
the corresponding values at neighboring sites.
This is well justified for a sufficiently shallow trapping potential. 
In this approximation, the ground state energy reads
$E=\langle\psi_0 |\hat{H}_0| \psi_0 \rangle+\Delta E_B+\Delta E_F+O({J}^3)$, where
$\Delta E_B=
2{J}d\sum_i\left(
|\psi_B^i|^2(
	1+{J} d r_i
)
\right)$,
with
$r_i=
(4\bar n_i + 2c_i+2)/(c_i^2-1)$,
$c_i=1-2\bar n_i-{V}_i+{\mu}_B-{U}_{BF}\bar m_i $,
and $d$ is the coordination number
of a $D$-dimensional cubic lattice ($d=6$ in three dimensions). 
We are now in the position to apply the Landau argument 
to determine the phase boundaries within LDA. 
If $1 +J d r_B^i> 0$, then the
approximate energy functional 
is minimized by having $| \psi_B^i| =0$, which
corresponds to the incompressible 
Mott situation for the bosons. In turn, for $1+J d r_B^i < 0$ 
the minimization requires $| \psi_B^i| >0$, and the bosons are superfluid.
Exploiting this property, we can determine the phase boundary 
between the hopping-dominated
and the Mott regime at each site, corresponding to
$J= -1/(d r_i)$.
To find the boundaries for the fermions, we consider
that for small $J$ and within LDA the 
bosons alter the fermionic chemical potential, introducing
an effective site dependent chemical potential
$\bar{\mu}_F^i=\mu_F-U_{BF}\bar{n}_i-V_i$. At each site $i$ we 
then consider the corresponding (infinite) homogeneous problem
$\hat{H}_F^i=-J
	\sum_{\langle l,j\rangle}
	(
	\hat{f}_l^\dagger\hat{f}_j +
	\hat{f}_j^\dagger\hat{f}_l )
	-\bar{\mu}_F^i\sum_l\hat{n}_F^l$,
which is appropriate for sufficiently shallow external potentials. 
This Hamiltonian is diagonal in Fourier space, so that the 
exact spectrum is given by
$
\varepsilon_k=-\bar{\mu}_F^i-4J\sum_{\delta=1}^D\cos(k_\delta),
$
where $k=(k_1,...,k_D)$, the lattice
spacing being set to $1$ without loss of generality.
Therefore, when $-\bar{\mu}_F^i-2dJ > 0$, the ground state has no
fermions present, being obviously a Mott state.
Similarly, for $-\bar{\mu}_F^i+2dJ < 0$ the ground state is
a Mott state with exactly one fermion at each site.
Fig.\ \ref{2} shows the phase regions for 
an inhomogeneous Bose-Fermi mixture in a three-dimensional
lattice with a weakly confining parabolic potential.
The solid lines depict the boundaries between 
Mott and hopping dominated regions, respectively, as evaluated using
the above approach. Not surprisingly, one observes that at the center of the 
trap, where the potential acquires its minimum, 
lower values of the hopping are needed for the transition to
the hopping-dominated regime. 
For appropriate fixed $J$, different spatial domains develop from the
center of the trap. Depending on the value of $\mu_B$, one observes 
an alternating sequence of Mott and hopping-dominated
domains.
%
An important new feature that emerges in inhomogeneous BFH systems
differing from the situation encountered in pure bosonic
or fermionic systems is a modulation of the phase
regions due to the boson-fermion interaction. 
This can be understood by comparing the phase boundaries for the
interacting mixture with the non-interacting case $U_{BF}=0$. 
The boundaries are represented as dashed lines in Fig.\ \ref{2}. 
For the chosen parameters, the presence
of the fermions in the center of the trap is reflected by a 
tendency to form Mott domains for bosons. 
Comparing this functional behavior with the 
fermion number per site in the case of vanishing hopping as depicted
in Fig.\ \ref{1} we see that the state diagram for the
bosons is modified when the fermion number per site
is exactly one. In turn, the presence of the bosons heavily modifies the
boundaries between the Mott and the hopping-dominated domains for the
fermions: the hopping-dominated regions are pushed outwards,
and the value of the integer boson occupation number per site 
in the Mott phase sets the scale of this phenomenon. 

{\it III) Finite hopping: variational theory.} --
In Fig.\ \ref{2} 
we have also represented the variance of the on-site 
densities 
$\sigma^i_{B/F}$.
They are determined using the following variational approach. 
We consider at each bulk site $i$ the
corresponding infinite homogeneous lattice Hamiltonian, $\hat{H}_i$.
The minimization of $\langle \phi_i|\hat{H}_i|\phi_i\rangle$
over all state vectors will be replaced by a minimization over state vectors
respecting the univalence superselection rule,
$|\phi_i\rangle=|\phi_B^i\rangle|\phi_F^i\rangle$.
For the bosonic sector we introduce a Gutzwiller-type
ansatz, $|\phi_B^i\rangle=\prod_l\sum_{n_l}b^{i}_{ n_l}
|n_l\rangle$
(see, e.g., Refs.\ \cite{Krauth,Jaksch}
and references therein),
where the $b^i_{n_l}$ form a probability 
distribution at each site $l$, $n_l=0,1,...$ .
After an exact discrete Fourier transformation of the fermionic
operators, $\hat{f}_l=\frac{1}{\sqrt{V}}\sum_k\hat{a}_k\me^{\mi kr_l}$,
we have for each site $i$,
\begin{eqnarray*}
\langle \phi_i|\hat{H}_i|\phi_i\rangle
	=
E_B^i+\sum_k\varepsilon_k^i\langle\phi_F^i|\hat{a}_k^\dagger\hat{a}_k|\phi_F^i\rangle,\hspace{2cm}&&\\
	E_B^i=-J\sum_{\langle l,j\rangle}\langle\phi_B^i|\hat{b}_l^\dagger\hat{b}_j
	+ \hat{b}_j^\dagger\hat{b}_l|\phi_B^i\rangle \hspace{3cm}&& \nonumber \\
	+\sum_l\langle\phi_B^i|\hat{n}^l_B(\hat{n}^l_B-1)|\phi_B^i\rangle
	+(V_i-\mu_B)\sum_l\langle\phi_B^i|\hat{n}^l_B|\phi_B^i\rangle,&&
\end{eqnarray*}
with
$\varepsilon_k^i=-4J\sum_{\delta=1}^D\cos(k_\delta)-\mu_F
-U_{BF}\langle\phi_B|\hat{n}_B^l|\phi_B\rangle-V_i$. Therefore, the
state vector 
$|\phi_F^0\rangle=\prod_{k,\varepsilon_k^i<0}\hat{a}_k^\dagger|0\rangle$
minimizes the energy expectation value at fixed Gutzwiller amplitudes,
\begin{eqnarray*}
E_{\text{min}}^i(b_0^i,b_1^i, \cdots)=
E_B^i+\sum_{k, \varepsilon_k^i<0}\varepsilon_k^i.
\end{eqnarray*}
To determine the ground state, we
have to minimize $E_{\text{min}}^i$ at each site $i$. Because this energy functional
is not convex, the energy landscape exhibits local minima and 
determining the ground state 
leads to a non-convex optimization problem.
However, the problem can be solved numerically
using a simulated annealing method \cite{annealing}. The regions
with exactly vanishing 
local variance, $\sigma^i_{B/F}$,
identify the respective Mott regions 
(dark blue/grey in Fig.\ \ref{2}).
Qualitatively, we obtain 
very similar results in the perturbative and in the
variational treatments.
The perturbative findings are valid for small hopping only, while
the numerical analysis relies on the Gutzwiller ansatz for bosons,
which is appropriate in high spatial dimensions ($D=3$) and in the
superfluid regime \cite{Variational,Approx}.
The behavior shown in Fig.\ \ref{2} is thus 
a genuine effect of the boson-fermion
interaction in the mixture, as it is predicted by
the considered approximation schemes.
For a system of harmonically trapped bosons it has been shown that
the appearance of a Mott-insulator domain within a shell of
superfluid atoms leads to satellite peaks in the global momentum
distribution \cite{Experiment}.
This feature is accessible in experiments and can in particular be used as
an indication for the effect of the fermions on the boson Mott transition.

\begin{figure}
\includegraphics[width=\columnwidth]{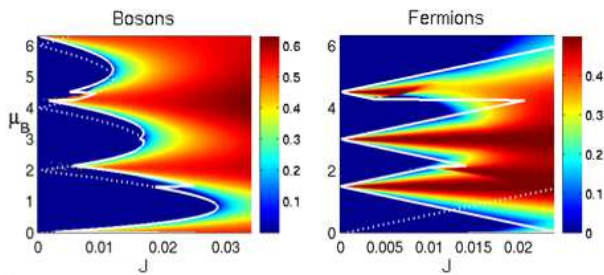}
\caption{State diagram for central sites of 
bosons (left) and fermions (right) for $U_{BF}=0.3$ and $\mu_F=\mu_B/5$.
Depicted are the phase boundaries as determined using  
the argument from section II in LDA (solid lines) and 
in the Gutzwiller variational theory (color encoding). The dashed lines
correspond to boundaries for
$U_{BF}=0$, determined
using the Landau argument from section II for the bosons, and exactly for 
the fermions.}\label{3}
\end{figure}

{\it IV) Behavior at the center of the trap: bulk properties.} --
For the central sites, within LDA, the inhomogeneous case
is equivalent to the homogeneous case.
To interpret the findings, we first recall how the
phase diagram for the fermions would look like in 
the homogeneous case in the
absence of bosons. In this case the BFH model
describes
a spinless fermion system with hopping contributions only.
It can be solved without approximation
as before with the help of a discrete 
Fourier transformation. The Mott states with
exactly one or zero fermions per site can be distinguished
from the hopping-dominated states,
yielding a linear behavior of the phase boundary as a function of 
$J=J_F$ (see section II).
This is depicted in Fig.\ \ref{3} with a dashed line.
Within the perturbative treatment, the effect of the bosons is to
give rise to an effective fermionic chemical potential, 
reflecting the change of the number of bosons per site in the Mott
phase. This in turn leads to integer discontinuous jumps in the phase boundaries. 
In this way, the presence of the bosons modifies the fermionic phase
diagram. In turn, the 
presence of the fermions modulates the phase diagram for the bosons
as compared to the standard mean-field phase diagram of
the Bose-Hubbard-model. Notably, the lobes associated to
different boson numbers per site in the Mott insulator do not 
necessarily touch the straight line corresponding to $J=0$.  
Again, we have compared these findings with the results obtained
from the numerical analysis introduced in section III.
The general behavior of the regions with  exactly vanishing
density variance within Gutzwiller approximation, 
and Mott regions according to the perturbative results
is very similar. However, the discontinuities are less 
pronounced within the variational approximation. 
This is due to the fact that 
in perturbation theory the zeroth order
contribution is manifestly discontinuous.
We have compared this behavior with the results obtained from an exact
diagonalization of the Hamiltonian for small systems
\cite{Remark}, obtaining qualitatively
identical conclusions.

In conclusion, we have studied in detail the phase structure
of the ground state of trapped 
inhomogeneous Bose-Fermi mixtures in optical
lattices. 
The inhomogeneity leads to domains of Mott plateaux
and hopping-dominated regions, where a complex interplay
between interacting bosons and fermions is displayed.
Introducing a new numerical method that treats fermions
without approximation, we were able to visualize for
the first time the effects of this complex interplay on
the domain structure of both species and
present the phase diagrams for the homogeneous case.
These results will be compared with DMRG-methods for
one-dimensional lattices in forthcoming work.
The findings reported in the present work should provide
a guideline and should be amenable to direct testing in 
the upcoming experiments \cite{Lattice} with trapped mixtures 
of bosonic and fermionic atoms in optical lattices.

We warmly thank A.\ Albus for constructive
comments and key input
in earlier stages of this project, and 
M.\ Baranov,
H.\ Fehrmann,
M.\ Fleischhauer, 
M.\ Lewenstein, 
L.\ Plimak,
and M.\ Wilkens for discussions.
This work was supported by the EU
(ACQP, QUPRODIS, QUIPROCONE), the ESF (project BEC2000+),
the DFG (SPP 1116, SPP 1078), the INFM,
INFN, and MIUR under the 
project PRIN-COFIN 2002.


\begin{thebibliography}{99}
\bibitem{single-band}
	Necessary conditions for the single-band approximation
	to be valid are a sufficiently low temperature and a fermionic filling 
	factor smaller than one. Furthermore, the lowest band needs to be
	well separated from the other bands, which in optical lattices
	can be realized by sufficiently deep lattices,
	 and the interaction energies
	must be sufficiently small.
\bibitem{Approx}

	For purely bosonic systems, we were able
	to reproduce the findings for the local density of Ref.\ \cite{Batrouni}
	without noticeable
	difference, providing an indication for the
	validity of the LDA.
\bibitem{Remark}
	This diagonalization has been performed for $8$	lattice sites, $5$ fermions and $3$ bosons, employing
	Arnoldi techniques.
\end{thebibliography}
\end{document}